\documentclass[12pt,osajnl2]{revtex4}
\usepackage{color}

\begin{document}

\title{Optical control of diffuse light storage in an ultracold atomic gas }

\author{Leonid V. Gerasimov,$^{1}$ Igor M. Sokolov,$^{1,2}$ Dmitriy V. Kupriyanov,$^{1}$
\\ Rocio G. Olave,$^{3}$ and  Mark D. Havey$^{3,*}$}
\address{$^1$Department of Theoretical Physics, State
Polytechnic University,\\ 195251, St.-Petersburg, Russia}
\address{$^{2}$\em{Institute for Analytical Instrumentation, Russian
Academy of Sciences, 198103, St.-Petersburg, Russia}}
\address{$^3$Department of Physics, Old Dominion University, Norfolk, VA
23529, USA}
\address{$^*$Corresponding author: mhavey@odu.edu}

\begin{abstract}
We show that coherent multiple light scattering, or diffuse light
propagation, in a disordered atomic medium, prepared at ultra-low
temperatures, can be be effectively delayed in the presence of a
strong control field initiating a stimulated Raman process. On a
relatively short time scale, when the atomic system can preserve
its configuration and effects of atomic motion can be ignored, the
scattered signal pulse, diffusely propagating via multiple
coherent scattering through the medium, can be stored in the spin
subsystem through its stimulated Raman-type conversion into spin
coherence. We demonstrate how this mechanism, potentially
interesting for developing quantum memories, would work for the
example of a coherent light pulse propagating through an
alkali-metal atomic vapor under typical conditions attainable in
experiments with ultracold atoms.
\end{abstract}

\ocis{020.1670, 270.1670, 270.5565, 270.5585}

\maketitle

\section{Introduction}
\noindent
Organization of controllable and lossless light transport through a bulk medium
is an important task for applications in quantum information science and for
implementation of quantum memories in particular, see recent reviews
\cite{HammSorPol,QMemories} and reference therein. The crucial problem is in
identifying those physical systems where transport of a signal light pulse can
be reliably delayed, safely stored and retrieved on demand. There are several
physical schemes which are considered as candidates for effective light
storage. A long-lived quantum memory for a period of a few milliseconds has
been recently demonstrated in \cite{ZDJCMKK} via Raman scattering on atomic
lattice. However because of relatively low overall efficiency such a long-lived
atomic memory could be mainly applied to realization of a deterministic single
photon source. Other schemes, more oriented toward the problem of a quantum repeater,
are based on the $\Lambda$-type conversion of the signal pulse, propagating
through a dense atomic system, into the non-decaying spin coherence. That
could be done under conditions of either electromagnetically induced transparency
(EIT), or a stimulated Raman process, or via a photon echo induced by either
controllable reverse inhomogeneous broadening (CRIB) or an atomic frequency comb
(AFC). The EIT based scheme in a hot atomic ensemble was successfully
demonstrated in \cite{NGPSLW} with an efficiency of around 40\% and also in
\cite{CDLK}, where a single photon state was stored in two ensembles of cold
atoms with an efficiency of 17\%. The echo scheme is preferably implemented in the
solid-state systems and proposed to be effective in a multimode
configuration. As has been recently reported in \cite{AWSHKAULSRG} the quantum
memory based on AFC in an ion-doped crystal can approach an efficiency of around
35\%. At present, the number of reported experiments on atomic memories
is increasing rapidly; the reader can find representative references in recent
reviews \cite{HammSorPol,QMemories}.

However, despite obvious progress in developing quantum memories
during the last decade the highly effective and long-lived light
storage is still a challenging experimental task. For the long
distance quantum communication via a quantum repeater protocol
\cite{BDCZ,BouwmEkkertZeilinger} the requirements for the
reliability of the quantum memories are extremely robust. The
memory unit should be highly effective, noiseless and reproducible
on a controllable time scale at each segment of the multipath
quantum channel. The problem becomes even more subtle if it is
addressed to the quantum network operating with continuous
variables, see \cite{SSGLT-BG,BRPAS}. Any new physical approaches
would evidently be desirable for further progress in this
challenging and rapidly developing area of quantum information
science.

In the present article we establish that coherent multiple
scattering, or diffuse light propagation in a disordered atomic
medium prepared at ultra-low temperature, represents a promising
process for the effective light storage and organization of the
quantum memory protocol. We emphasis that the process as a whole
contains elements of coherent control of light scattering via
electromagnetically induced transparency and, at the same time,
relies on the coherence of multiple light scattering in an
ultracold atomic gas. The main point in this investigation is that
the $\emph{entirety}$ of a propagating signal pulse scattered by
the medium is not lost (as is normally estimated in standard
schemes based on either electromagnetically induced transparency
(EIT) or Raman process for forward propagating light), but
transformed into a diffusely propagating, but coherent, mode. This
mode can be effectively controlled and stored by initiating its
stimulated Raman-type conversion into atomic spin coherence. We
theoretically demonstrate how it can be done.

The paper is organized as follows. In section \ref{Sec2} we describe our basic
physical idea concerning how the light diffusion can be controlled with coherent light.
In this section we also detail our calculation approach. Section \ref{Sec3} presents the
results of our numerical simulations and gives the estimates of the delay
effect for the light scattered via the Rayleigh scattering channel. Section \ref{Sec4} focuses
on the problem of quantum memory and specifically we show one example of how the
Bell detection and the proposed quantum memory unit could be integrated in one
node of general quantum network. The main results of the paper are emphasized
in the conclusions.

\section{Light diffusion controlled via stimulated Raman process} \label{Sec2}

First consider the situation when a light pulse diffusely propagates through an
opaque atomic medium.  Here we will ignore in this process  effects associated
with atomic thermal motion. Such a situation may be approximately obtained for
light scattering by an atomic ensemble prepared in a magneto optical trap
(MOT)\cite{LightScatt_in_MOT}. Specifically, let us consider a ${}^{85}$Rb
sample and a pulse resonant with the closed $F_0=3\to F=4$ transition
associated with the $D_2$-line. As shown in Figure 1, the original
forward propagating light pulse is split by atomic scatterers into fragments
and emerges from the sample more or less isotropically. This process is normally
termed incoherent scattering, with the implied emphasis that coherence should be
associated with the forward transmitted part of the light. However, any light
fragment elastically scattered via a Rayleigh channel in any direction is
generated by a coherent response of the atomic polarization to the incident light.
Therefore the secondary and multiply scattered waves assemble coherently and
eventually create a configuration-dependent emergent spatial mode. Moreover, in
an optically dense configuration, only a negligible contribution of the input
light pulse is transmitted by the sample in the forward direction since the
free propagation length scale for a resonant photon $l_0\sim 1/n_0\sigma_0$,
where $n_0$ is the atomic density and $\sigma_0$ denotes the resonant light
scattering cross section, is much shorter than the sample length. The diffusion
mechanism dominates in dynamics of the light transport in such medium. Since
the upper hyperfine sublevels $F=3,2$ are far detuned from the selected
excitation channel, spontaneous Raman scattering makes a negligible contribution
to the diffusion process.

\begin{figure}[t]
\centerline{\includegraphics[width=8cm]{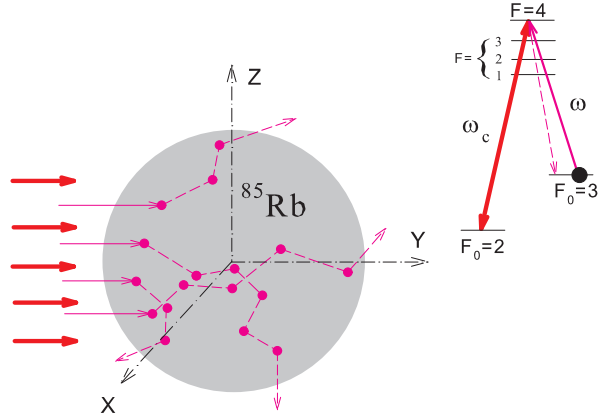}} \caption{(Color online)
Geometry of the proposed experiment and the excitation scheme, see text. The
diffuse propagation of a signal mode of frequency $\omega$ is indicated by
pink thin lines and arrows. The diffusion process is affected by a strong control
mode of frequency $\omega_c$ indicated by red and thick arrows.}
\label{fig1}
\end{figure}

Let us make a crucial modification of the above process  and
additionally apply a strong control mode between the lower $F_0=2$
and upper $F=4$ hyperfine sublevels. Of itself, this transition is
forbidden, but evidently the applied field opens the stimulated
Raman channel via far-off-resonance coupling with other hyperfine
sublevels of the upper state $F=3,2$. This consists of many
Lambda-type couplings strongly interfering in the entire process,
see Appendix \ref{A} for details. Now both the elastic Rayleigh
and stimulated Raman processes affect transport of the coherent
signal pulse. The influence of Raman scattering is much weaker but
its associated transition probability accumulates with each step
of a multiple scattering path, that is, while the photon
propagates through the sample along a random and zigzag-type
diffusion path, see figure \ref{fig1}. The crucial property of the
stimulated Raman interaction is that the photon scattered into the
control mode will be, with great probability, retrieved into the
signal mode and we shall discuss how this process affect the
signal pulse propagation in more detail.

We restrict our consideration by tracing the dynamics of a weak signal mode
when its outgoing amplitude is obtained by linear transformation of its
original amplitude in the incident wave. Then the light diffusion can be
analyzed via following the transformation of only the
first order correlation function of the signal light. This correlation
function, being the function of spatial coordinates $\mathbf{r}$ and
$\mathbf{r}'$ and time arguments $t$ and $t'$, is given by
\begin{equation}
D_{\mu\nu}^{(E)}(\mathbf{r},t;\mathbf{r}',t')=%
\left\langle E_{\nu}^{(-)}(\mathbf{r}',t')E_{\mu}^{(+)}(\mathbf{r},t)\right\rangle
\label{2.1}
\end{equation}
where $E_{\mu}^{(+)}(\mathbf{r},t)$ and $E_{\nu}^{(-)}(\mathbf{r}',t')$ are the
Heisenberg operators of positive and negative frequency components of the
electric field, $\mu$ and $\nu$ are vector indices in the plane orthogonal to
the light local propagation direction, and the angle brackets denote the
statistical averaging over the initial state of the light and atoms. The total
intensity of the scattered light $I(t)$ is expressed by the sum of diagonal
components of the correlation function considered at coincident moments of time
and integrated over all propagation directions outside the medium.

In a dilute medium, when the mean transport path is long enough such that
$l_0\gg k^{-1}=\lambdabar$, for a light ray propagating in any direction there
is a preferable coherent enhancement for its current forward propagation. This
means that along any ray, for a short mesoscopic scale consisting of a large
number of atoms, there is only a slight attenuation of the propagating wave.
Such an attenuation comes from the events of random single atom scattering (as
was commented above, usually termed incoherent scattering), which have small
but not negligible probability. The mesoscopically averaged light propagation
is properly described by the macroscopic Maxwell equations and characterized by
the sample susceptibility tensor
\begin{equation}
\chi_{ij}(\omega)=\chi_{0}(\omega)\delta_{ij}+\chi^{(AT)}_{ij}(\omega),\ \ \ i,j=(x,y,z)%
\label{2.2}
\end{equation}
which, in our case, consists of a dominant linear isotropic term
$\chi_{0}(\omega)$ and a nonlinear anisotropic correction
$\chi^{(AT)}_{ij}(\omega)$ associated with the "dressing" of the upper state by
the control mode, which is known as the Autler-Townes (AT) effect
\cite{AutlerTownes,LethChebt}. The single atom scattering is described by the
scattering tensor, which can be qualitatively interpreted as a microscopic
susceptibility related to a local inhomogeneity associated with an isolated
atomic scatterer. Both the characteristics susceptibility and scattering tensor
are subsequently introduced in Appendix \ref{A}.

The perturbation theory diagram analysis shows that under the discussed conditions
the correlation function of the outgoing probe pulse can be evaluated via
summing of so-called "ladder"-type diagrams. The "crossed"-type diagram
correction should be added only for evaluation of the backwards scattered radiation
\cite{LightScatt_in_MOT}, which makes, for the conditions considered here, a negligible contribution to the light
scattered into the total solid angle. After mesoscopic averaging, the actual
light wave in the sample can be visualized as a set of unknown zigzag paths,
whose vertices consist of atoms scattering the light from the direction of
forward propagation, see Figure \ref{fig1}. Any randomly chosen path contains a
chain of atoms located at the vertices and their number is just associated with
the scattering order. Such a qualitative physical picture correctly reproduces
the entire calculation procedure based on the diagram analysis and on the
Monte-Carlo simulation scheme. Each chain of atomic scatterers makes a partial
contribution to the formation of the outgoing wave and the total intensity
$I(t)$ of the probe pulse emerging the sample can be expanded in the following
series
\begin{equation}
I(t)=\sum_{n=0}^{\infty}I^{(n)}(t)
\label{2.3}
\end{equation}
where each partial contribution $I^{(n)}(t)$ relates to the $n$-th order of
multiple scattering. It is important that, as a result of mesoscopic averaging,
this series is converging rapidly and only the multiple scattering of
relatively low orders (around squared optical depth of the sample) contribute
significantly to formation of the pulse intensity or its correlation function
(\ref{2.1}).

\begin{figure}[t]
\centerline{\includegraphics[width=8cm]{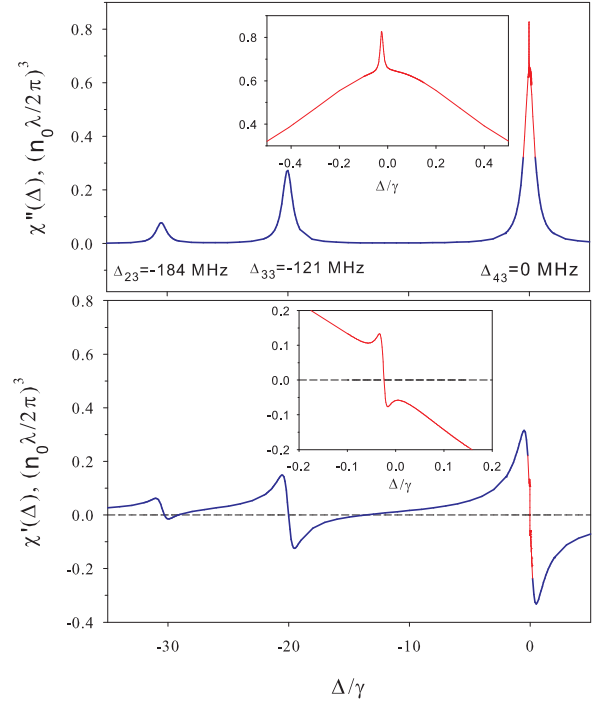}}
\caption{(Color online) The dielectric susceptibility of the
sample $\chi(\Delta)=\chi'(\Delta)+i\chi''(\Delta)$ as a function
of frequency detuning $\Delta=\omega-\omega_{43}$ in units of
$n_0\lambdabar^3=n_0(c/\omega)^3$. Upper and lower panels
respectively show the absorptive and dispersive components of the
susceptibility tensor in the case of linear and orthogonal
polarizations of the signal and control modes, see geometry
shown in panel (a) of Fig. \ref{fig3}.}
\label{fig2}
\end{figure}

In Figure \ref{fig2} we reproduce one example of our calculations of the sample
susceptibility for a geometry where signal and control modes are in linear but
orthogonal polarizations. As is clearly seen from the plotted graphs the dominant
contribution comes from the linear term and all three allowed resonance
transitions for excitation from the lower $F_0=3$ to the upper $F=4,3,2$
hyperfine sublevels of $D_2$-line are shown. These resonances have Lorentzian
profiles characterized by the atomic natural line width $\gamma$. The nonlinear
susceptibility of the medium manifests itself as a narrow but large amplitude
spectral feature in the "absorption" (actually associated with the scattering
losses) and dispersion profiles near the $F_0=3\to F=4$ resonance transition.
For the calculated AT resonance the amplitude of the control mode is given by
the Rabi frequency $\Omega_c=2|V_{nm'}|=3\gamma$, which was chosen for a
particular transition between the states $|m'\rangle=|F_0=2,M_0=2\rangle$ and
$|n\rangle=|F=3,M=2\rangle$, see Appendix \ref{A}. To tune the AT resonance nearer
the maximum of the linear "absorption" Lorentzian profile the frequency of the
control field was slightly shifted from the frequency $\omega_{42}$ such that
$\omega_c=\omega_{42}-0.4\gamma$.

The dispersive part of the nonlinear susceptibility governs the
stimulated Raman process, which is responsible for an additional
delay in the diffuse propagation of the signal pulse. If the pulse
spectrum is chosen to be spectrally broader than the AT peak, such
that spontaneous Raman scattering is negligible, but still
narrower than $\gamma$, then the diffusely propagating light can
be effectively delayed for a period of time comparable to or
longer than the pulse duration. Roughly speaking, the bandwidth of
the AT resonance is scaled as
$\Gamma_{\mathrm{AT}}\sim\Omega_c^2\gamma/\Delta_{\mathrm{hpf}}^2$,
where $\Delta_{\mathrm{hpf}}$ is the hyperfine splitting in the
upper state, such that the required condition
$\Gamma_{\mathrm{AT}}\ll\gamma$ is easy to fulfill. In each
partial intensity $I^{(n)}(t)$ the electric field amplitude can be
approximately expressed by a convolution transform of
non-disturbed field, which would propagate in the medium without
control mode, and the transfer-type function associated with the
nonlinear term in Eq.(\ref{2.2}) and expressing the action of the
control mode. This action consists of the following; along each
multiple scattering path the nonlinear coupling with the control
mode affects the pulse propagation via its stimulated Raman-type
conversion into the spin coherence in the atomic subsystem and
recovering into the signal mode with delay. Evidently this effect
being applied to the diffusely propagating light would be more
effective than in the standard one dimensional geometry associated
with only forward propagation. If the control mode is then
switched off, this diffusely propagating signal pulse would be
stored in the spin coherence. The process of light transport can
be regenerated if the control mode is switched on again after a
controllable delay time.

\section{Numerical results} \label{Sec3}

\begin{figure}[t]
\centerline{\includegraphics[width=17cm]{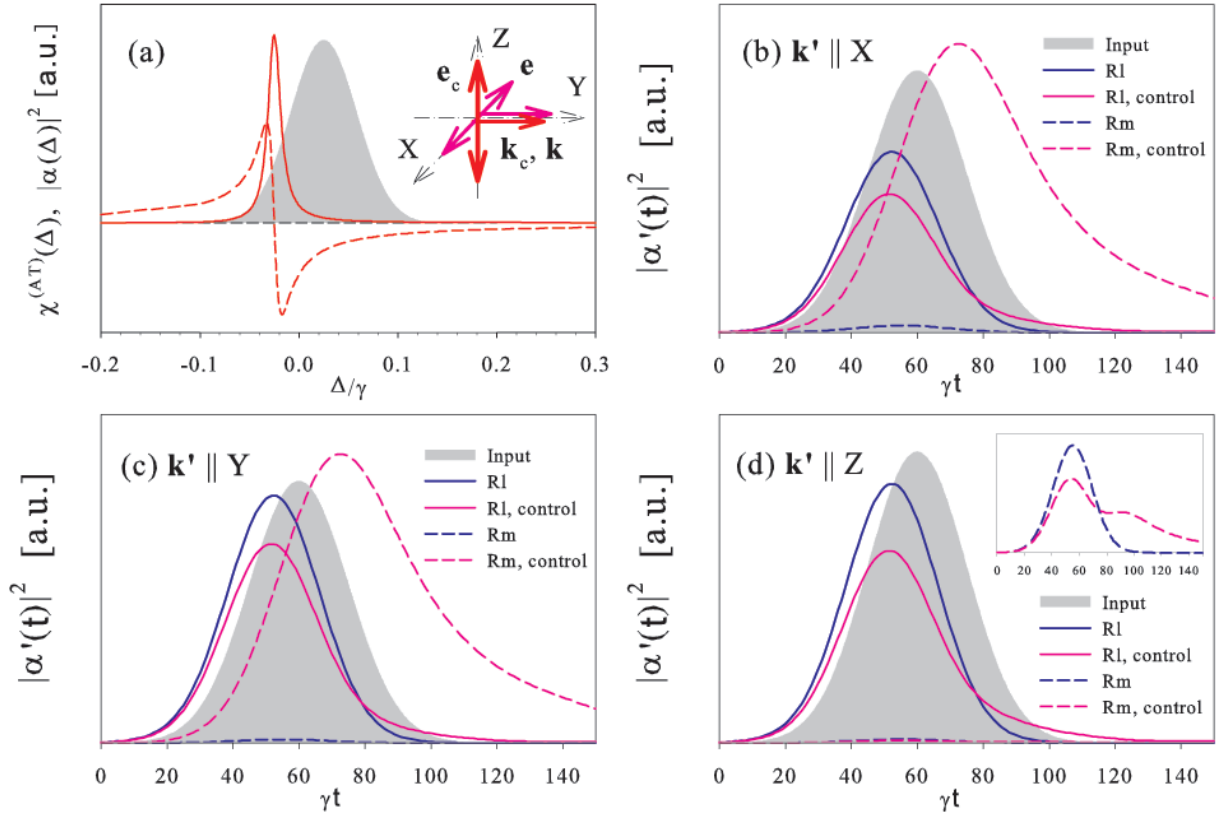}}
\caption{(Color online) (a): Pulse spectrum vs. Autler-Townes
resonance and polarization directions for the control
$\mathbf{e}_c$ and signal $\mathbf{e}$ modes propagating along the
$Y$ direction, see Fig. \ref{fig1}. (b)-(d): Fragment of the
signal pulse $\alpha'(t)$ with wave vector $\mathbf{k}'$ generated
by single scattering from the central point of the cloud in the
directions $X$ (b), $Y$ (c), and $Z$ (d). The blue curves indicate
the effect of elastic Rayleigh (Rl, solid) and Raman (Rm, dashed)
scattering without the action of the control mode.  The pink
curves show the corresponding scattering and delay in the presence
of the control mode. Comparison of the reference input profile
(shadowed) and the scattered light shows superluminal propagation
of the scattered light in the spectral domain of anomalous
dispersion, see Fig. 2. The delay manifests itself in the long
time asymptote of the pulse tail.} \label{fig3}
\end{figure}

\noindent The complete multiple scattering analysis, based on
Monte-Carlo numerical simulations, which is evidently required for
the protocol demonstration, will be done later and published
elsewhere. Here we illustrate the main effects by our calculations
done for a single scattering event at the central point of the
cloud and at the observation angles coinciding with or orthogonal
to the propagation direction of the incident pulse, see geometry
and reference frame shown in Figure \ref{fig1}. In our
calculations we have assumed a sample generated in a typical MOT
configuration characterized by a spherical Gaussian atom
distribution with a square variance $r_0^2$ and peak density $n_0$
such that the local density at spatial point $\mathbf{r}$ is given
by
\begin{equation}
n(\mathbf{r})=n_0\exp\left(-\frac{\mathbf{r}^2}{2r^2_0}\right)%
\label{3.1}
\end{equation}
The optical depth $b_0=\sqrt{2\pi}\sigma_0n_0r_0$, associated with the customary
Beer's law definition, was taken as $b_0\sim 10$, which
is readily attainable in our laboratory setup. The normalized slow-varying
amplitude of the signal pulse $\alpha(t)$ has a Gaussian profile, which
spectrum $\alpha(\Delta)$ compared with the AT resonance is shown in panel (a)
of figure \ref{fig3}. In the time domain the pulse is parameterized as follows
\begin{equation}
\alpha(t)=\frac{2}{\left(2\pi T^2\right)^{1/4}}\exp\left[-i\Delta t-4\frac{(t-T)^2}{T^2}\right]%
\label{3.2}
\end{equation}
where $\Delta=\omega-\omega_{43}=0.025\gamma$ is the small shift
between the pulse carrier frequency and the frequency of the
atomic $F_0=3\to F=4$ transition, see Figure \ref{fig1}. This
shift allows to optimize the overlap between dispersion part of
the AT resonance, calculated in previous section, and the pulse
spectrum, see Figure \ref{fig3}(a). The pulse parameters are
specified by the mean arrival time $T$ and by the pulse duration
(also $\sim T$) and determined by one parameter $T=60\gamma^{-1}$.
The spatial distribution of the incoming pulse is considered as
flat with a radial aperture $0.5r_0$.

Panels (b),(c), and (d) of Figure \ref{fig3} respectively show the
single scattering response from the central point of the atomic
cloud (i.e. a single scatterer response) in the direction of $X,Y$
and $Z$ axes such that the optical path in the elastic channel is
equal to $b_0$ for all the plotted graphs. We consider both the
elastic and inelastic scattering channels and extend the
definition of Rayleigh scattering by also including the
contribution of elastic Raman process in it. It seems natural for
our discussion to refer to the entire elastic channel as Rayleigh
and inelastic as Raman since all the light scattered elastically
undergoes diffusive transport and can be delayed and stored under
the protocol described in the previous section. The stimulated
Raman process governing this protocol, affects the system
dynamically and a certain leading edge of the light pulse will
leak because in the beginning stage this dynamical action is
always weak. The delay effect is seen most clearly in the long
time asymptote of the pulse tale when the stimulated Raman
interaction has enough duration to be more fully manifested. This
effect is clearly seen in the plotted dependencies. The leakage
could be made smaller if the Rabi frequency of the control was
higher, but in this case more losses from spontaneous Raman
scattering would be expected. For each particular optical depth
$b_0$ there is an optimal balance between pulse duration and Rabi
frequency.

Even in an optimal situation the light scattered from the sample via
spontaneous (not stimulated) Raman process generates losses of the signal pulse
and induces noise (extra non-informative light) at the retrieval stage of the
protocol. As can be seen from the dependencies presented in Figure \ref{fig3},
in the absence of the control mode, the spontaneous Raman channel makes a fully
negligible contribution. To estimate the relative amount of the Raman
contribution in the single scattering event one should pay attention that for
the presented data, in accordance with the Beer's law, the intensity of the
Rayleigh fraction is attenuated by the factor $e^{-5}\sim 0.007$ from its
amount at the central point of the atomic cloud where it was generated. But as
follows from the presented data the Raman fraction, which is not attenuated and
propagates to the detector freely, reveals a much smaller contribution than the
Rayleigh part. In the presence of the control mode the Raman fraction is
significantly enhanced for the scattering in the directions of $X$ and $Y$ axes
and becomes even greater than the Rayleigh part. That is a direct consequence
of the AT effect and caused by the essential overlap of the pulse spectrum with
the "absorption" part of the AT resonance, see Figure \ref{fig3}(a). But again,
if one takes into account that the Rayleigh contribution is actually $e^{5}\sim
150$ times greater, then it will become clear that the losses via the
spontaneous Raman channel represents still a negligible effect.

As was indicated in the beginning of this section, the relative
amount of spontaneous Raman losses would be most realistically
estimated via a complete Monte-Carlo simulation for the scattering
into the overall solid angle. That would be given by the sum
(\ref{2.3}) and include all orders of multiple scattering. But let
us point out that the spontaneous contribution is expected to be
generally small because of the small overlap of the pulse spectrum
with the AT resonance, as is is shown in the first panel of Figure
\ref{fig3}. In an ideal situation we expect that there would be
low and negligible probability for such an incoherent transition
initiated by spontaneous Raman scattering.

It is important that the delay effect, for light emerging the
sample via the Rayleigh channel, can be observed from the
dependencies of Figure \ref{fig3}. The incoming pulse, with an
intensity reduced by a factor $3\cdot 10^{-10}$ (recall that the
plotted data relate to the contribution of a single atomic
scatterer), is shown here as a shadowed reference profile. In the
case of only a single scattering event the delay of the scattered
fragment, shown in panels (b)-(d) of Figure \ref{fig3}, is
partially masked by a superluminal effect caused by anomalous
dispersion, but clearly seen in the long time asymptote of the
pulse tail. Let us stress that in these graphs the delay should be
associated with the scattered fragment and not referred to the
original pulse. The superluminal effect is only a manifestation of
the fact that the leading edge of the incident pulse mainly
contributes to the single scattering signal. The main important
point is that if we accumulate all the contributions of a multiple
scattering chain (see Fig. \ref{fig1}), the delay can be enhanced
by the factor of $b_0$ while scaling the length of a real random
path for the signal photon in units of its mean straight line path
$b_0l_0$. We emphasize here that the contribution of the Raman
channel is expected to be small in each scattering order. For some
scattering angles it is even negligible, as is shown in the inset
of panel (d) of Figure \ref{fig3} for scattering in the
$Z$-direction, where the Raman signal (not attenuated) has been
increased by a scale factor of 100.

\section{Application to the quantum memory} \label{Sec4}

\noindent Let us briefly discuss how the results of the previous sections could
be applied to the quantum memory problem. The main inconvenience of the
proposed scheme is that it transforms the signal pulse into its scattered
fragments distributed over a substantial solid angle. However, we emphasize
that this should not be considered a fundamental barrier for further
applicability of the discussed mechanism in a quantum memory protocol. In
principle the light scattered by a small macroscopic object like an atomic
cloud in our case, which has a spatial scale around  $1\,\mathrm{mm}$ or less,
can be accumulated and focused in a small image spot with an ideal optical
system and then detected, see Ref.\cite{BornWolfMandel}. In the linear optics,
operating with mirrors, lenses etc., the image created by the retrieved signal
pulse, would have a spatial profile and polarization, which would be unique for
each particular mode of the incoming signal pulse but differ from it. Evidently
there are some losses of the pulse intensity in this process and additional
noise light, coming from other scattering channels, could contribute to the
observation channel. We anticipate that, for linear optics in some cases, this
noise can be eliminated in the observation channel after a round of
optimization. In this sense we would like to point out one recent proposal of
the effective and noiseless conversion of a plane to spherical wave with
parabolic mirror \cite{SMKLPL}.

In an alternate possibility to be explored, recovery of the scattered light
over a large solid angle may be overcome by applying a time reversal approach.
The spin coherence, which is created by diffusely propagating light, reveals a
so-called quantum hologram of the scattered light. If the hologram was
transformed via a non-linear complex conjugate operation, then one could
recover the object/signal pulse in its original profile, i.e. its spatial mode
and polarization. Approximately this could be done if we set the entire process
to evolve in the spatial area bounded by a non-linearly reflecting medium known
as a phase conjugate mirror, see \cite{BornWolfMandel}, the outgoing signal
light would be a time reversed copy of the incident pulse. However for such a
mirror there is always extra radiation (noise) created by the mirror itself. In
the ideal conditions the expected relation between the average photon number in
the outgoing pulse and the similar photon number in the incoming pulse is given
by $\bar{n}^{(\mathrm{out})}=\bar{n}^{(\mathrm{in})} + 1$. The additional
photon arrives because of noise and it cannot be eliminated in the observation
channel via any optimization procedure.

\begin{figure}[t]
\centerline{\includegraphics[width=8cm]{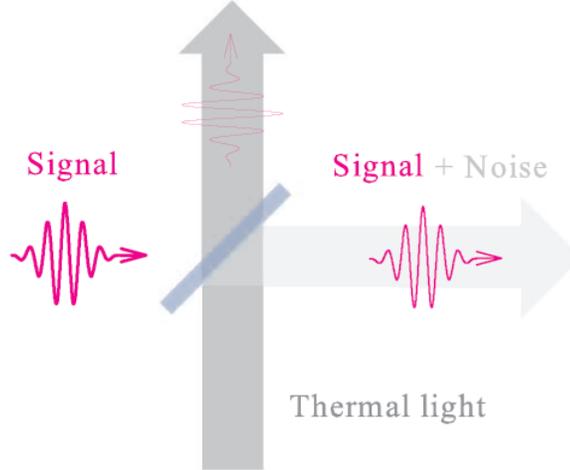}}
\caption{(Color online) Equivalent scheme for the quantum memory
channel. The signal single photon pulse passes through a
beamsplitter, with transmittance equal to the efficiency of the
memory channel. The additional noise is admixed via sending the
thermal light onto the second port of the beamsplitter.}
\label{fig4}
\end{figure}

In the single photon quantum memory protocol both the observation schemes can
be considered as equivalent to a beamsplitter, which transmission efficiency
$\eta$ is equal to the overall efficiency for write-in and readout stages of
the protocol and also includes technical losses of the detection channel, see
Figure \ref{fig4}. The noise, associated with other possible scattering
channels or radiation sources, can be introduced in this scheme via sending the
thermal light onto the second input port of the beamsplitter, such that it
would generate the Gaussian (also thermal) light with average number of photons
$\bar{n}$ in the observation channel. The observation channel in this model is
performed by the output port of the beamsplitter, see Figure \ref{fig4}. The
single photon input state is described by the following Wigner function
\begin{equation}
W^{(1)}(\alpha)=\frac{8}{\pi}\left[|\alpha|^2-\frac{1}{4}\right]\,\mathrm{e}^{-2|\alpha|^2}%
\label{4.1}
\end{equation}
where $\alpha=x+ip$ is the complex field amplitude, and $x$ and $p$ are the
quadrature components of the signal mode. After mixing with the thermal light
the modified Wigner function in the observation channel is given by
\begin{equation}
W(\alpha)=\frac{4}{\pi(\bar{n}+1/2)}\left[\frac{\eta|\alpha|^2}{4(\bar{n}+1/2)^2}%
-\frac{1}{4}+\frac{\bar{n}+(1-\eta)/2}{2(\bar{n}+1/2)}\right]\,%
\exp\left[-\frac{|\alpha|^2}{\bar{n}+1/2}\right]%
\label{4.2}
\end{equation}
and includes losses with efficiency $\eta<1$ and admixture of thermal noise
with average number of thermal photon per mode $\bar{n}>0$. This Wigner
function generally describes many photon states and has non-vanishing overlap
with the vacuum state as well.

Let us simplify the expression (\ref{4.2}) and consider it in two limiting
situations: first, for a lossless channel ($\eta=1$) but with noise; and second,
with losses but without noise ($\bar{n}=0$). For the lossless configuration the
Wigner function (\ref{4.2}) describes the quantum state where the number of
photons per mode, which could be randomly detected in the observation channel,
runs the sequence $n=0,1,2,\ldots$. If we post-selected only single photon
events with $n=1$, then the probability to observe either signal or noise photon
would depend on $\bar{n}$ and $\bar{n}=1$ determines the balance situation,
when these probabilities are equal. This just relates to the ideal observation
conditions based on the phase conjugate mirror technique. If the quantum
information was encoded in the polarization state of the signal photon, when
the quantum statistical ensemble for the polarization state of the retrieved
photon would be described by the following Werner-type density matrix
\cite{Werner,BouwmEkkertZeilinger}
\begin{equation}
\hat{\rho}=x|\psi\rangle\langle|\psi|+\frac{1-x}{2}\hat{I}
\label{4.3}
\end{equation}
where $|\psi\rangle$ is the original polarization state and
$\hat{I}$ is a unit matrix in two dimensional Hilbert space. For
$\bar{n}=1$ the weighting parameter $x=1/2$ means equal admixture
of detecting events for either informative or noise photons. In
this case the achieved fidelity
$F=\langle\psi|\hat{\rho}|\psi\rangle=3/4$ is higher than
classical benchmark based on single measurement
$F_{\mathrm{meas}}=2/3$, see Ref.\cite{MassarPopescu}, but less
than optimal cloning benchmark $F_{\mathrm{clon}}=5/6$, see
Ref.\cite{BouwmEkkertZeilinger}. Evidently, to improve the quality
of the memory channel a certain distillation procedure is needed,
see Refs.\cite{HammSorPol,QMemories}, and we skip discussion of
this subtle problem here.

In the case of a noiseless channel ($\bar{n}=0$) the Wigner function (\ref{4.2})
performs the superposition of the single photon (with weight $\eta$) and vacuum
(with weight $1-\eta$) states of light. The post-selection of the single photon
outcome with probability $\eta$ makes a resource for the conditional quantum
memory scheme based on the consequence that each successful single photon
detection always gives us the informative photon. However, as was pointed
out above, this could be reliably done for detection of the retrieved photon with
the set of linear optical devices such that the original spatial mode and
polarization state of the signal photon would be unpredictably transformed by
the scattering channel. Then it might seem that such quantum memory would be
useless for the quantum information implication. But we shall show one example
where such a memory scheme could be applicable and could enhance reliability of
quantum network.

\begin{figure}[t]
\centerline{\includegraphics[width=8cm]{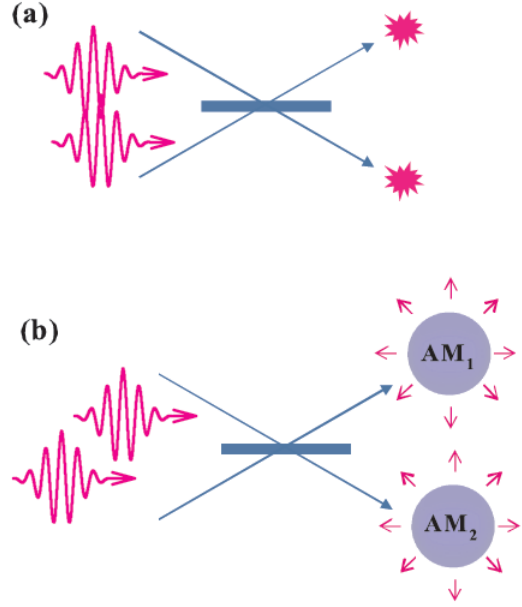}}
\caption{(Color online) Bell-type measurement based on
anti-bunching of the photon pair passed through a beamsplitter. In
the standard measurement scheme (a) both the photons should arrive
to the input ports simultaneously. With the quantum memories (b)
the measurement can be postponed such that the photons could
arrive at any time.} \label{fig5}
\end{figure}

In Figure \ref{fig5}(a) we reproduce the standard applied conditional
Bell-type projection measurement for a photon pair, considered as information
carrier, with a beamsplitter, see \cite{BouwmEkkertZeilinger}. In this
measurement if both the photons have arrived to the input ports of the
beamsplitter simultaneously then the detectors' clicks on each of two output
ports tells us that the photon state was projected onto the following
antisymmetric polarization state
\begin{equation}
|\Psi\rangle=\frac{1}{\sqrt{2}}\left[|\updownarrow,\leftrightarrow\rangle-%
|\leftrightarrow,\updownarrow,\rangle\right]
\label{4.4}
\end{equation}
where for the sake of concreteness we expanded this state in the basis of
linear polarizations. This can be also called as anti-bunching effect of the
photon pair passed through the beamsplitter.

The technical problem with such measurement is that the coherence volumes of
the incoming photons should be essentially overlapped i. e. the coincidence in
their arrival time is a crucial requirement for the measurement. If we set the
quantum memories in the output, how it is shown in Figure \ref{fig5}(b), this
inconvenience could be overcome. Indeed, the discussed memory protocol would
allow us to subsequently share the unknown state of each photon between two
atomic memory units $\mathrm{AM}_1$ and $\mathrm{AM}_2$. Then at the readout
stage both the photons could be released with much weaker control field than
that was used in the write-in stage. This operation leads to essential
extension of the outgoing pulse duration. As a consequence, the wavepackets of
the released photons would be overlapped and the photons would be
indistinguishable such that their further detection would give us the same
projection measurements onto the state (\ref{4.4}) as in the standard scheme.
This detection scheme would be insensitive to the spatial and polarization
modes of the outgoing photons and in case of general quantum network consisted
of many nodes the proposed integrated architecture of the Bell detector and
quantum memories could enhance the network reliability.

\section{Conclusion}

In conclusion, we have shown that a disordered atomic medium, which diffusely
scatters the light propagating through it, can be used as a memory unit for
light storage. The mechanism of the Raman-type conversion of the signal pulse
into the spin coherence is more effective in this case and the level of losses
is much smaller than for the reference process controlling only the forwardly
propagating part of the light.

Such effective light storage can be integrated into a quantum network in
combination with the standard conditional two-photon Bell detector. Bell
detection performs an anti-bunching of the photons overlapped with a
beamsplitter and then detected. The main advantage of the proposed scheme is
that with the quantum memories the anti-bunching of the detected photons can be
observed, even if the photons arrive to the input ports of the beamsplitter at
different times.

\acknowledgements This work was supported by the Russian Foundation for Basic
Research (Grant 10-02-00103 and the National
Science Foundation (NSF-PHY-0654226). L.G. acknowledges financial support from
the charity foundation $\textit{Dynasty}$.

\appendix
\section{The nonlinear susceptibility and scattering tensor} \label{A}

\begin{figure}[t]
\centerline{\includegraphics[width=8cm]{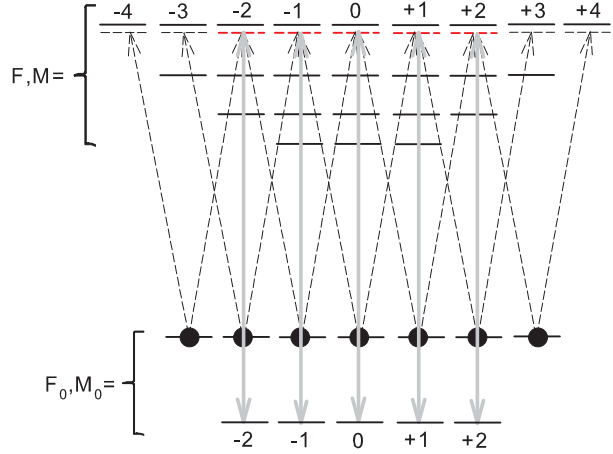}}
\caption{(Color online) The scheme of excitation of ${}^{85}$Rb, when the signal mode (dashed arrow lines) is
orthogonally polarized to the control mode (gray arrow lines). The control mode is linearly polarized and the
quantization axis $Z$ is directed along its polarization, see Figures \ref{fig1} and \ref{fig3}(a).}
\label{fig6}
\end{figure}

The susceptibility tensor responsible for elastic and mesoscopically averaged
propagation of the signal mode, or of any its scattered fragments, through the
atomic sample can be introduced in its major reference frame. For the
excitation geometry shown in Figure \ref{fig1} where the control mode is
linearly polarized along the $Z$-axis it has the following diagonal form
\begin{equation}
 \hat{\chi}=\left(\begin{array}{ccc}
 \chi_{\bot}&0&0\\
 0&\chi_{\bot}&0\\
 0&0&\chi_{\parallel}
 \end{array}\right)
\label{a.1}
 \end{equation}
where the $\chi_{\parallel}$ component determines the response of the atomic
polarization on its excitation by the probe mode linearly polarized along the
direction of the $Z$-axis and the transverse component $\chi_{\bot}$ gives the response
of the probe for any polarization in the $X,Y$-plane. As an example, in Figure
\ref{fig6} we show the excitation diagram of ${}^{85}$Rb when the polarization
vector of the signal mode is orthogonal to the polarization vector of the
control field.

For an ensemble of motionless atomic scatterers distributed in
space with density $n_0(\mathbf{r})$ the principal components of
the susceptibility tensor are given by
\begin{equation}
\chi_{\mathrm{j}}(\mathbf{r};\omega)=%
-n_0(\mathbf{r})\frac{1}{2F_0+1}\sum_{m}\sum_{n,n'}\frac{1}{\hbar}%
\left(\mathbf{d e_\mathrm{j}}\right)_{nm}^{*}%
\left(\mathbf{d e_\mathrm{j}}\right)_{n'm}\,%
G_{nn'}^{(--)}(\hbar\omega+E_m+i0)%
\label{a.2}%
\end{equation}
where $\mathrm{j}=X,Y(\bot);\ Z(\parallel)$ and $\left(\mathbf{d
e_\mathrm{j}}\right)_{nm}$ are the matrix elements between the ground
($|m\rangle$) and excited ($|n\rangle$ and $|n'\rangle$) states for the atomic
dipole operator projected on the major axis directions. The ground state Zeeman
sublevels are specified by the quantum numbers $m\equiv F_0,M_0$, where
$F_0=F_{+}\equiv I+1/2$, is the total (electronic $S=1/2$ and nuclear $I$)
angular momentum of the upper hyperfine sublevel and $M_0$ is its projection.
Each Zeeman sublevel has energy $E_m$ and all of them are assumed to be equally
populated. The quantum numbers $n\equiv F,M$ and $n'\equiv F',M'$ belong to the
set of those Zeeman states of the hyperfine sublevels $F=I+3/2,...,I-3/2$ of
the excited state in $D_2$ line, which are accessible by excitation for the
signal mode, see excitation diagram shown in figure \ref{fig6}.

The most important ingredient contributing to Eq.(\ref{a.2}) is the
retarded-type atomic Green's function of the excited state $G_{nn'}^{(--)}(E)$.
For the upper exited state $|n\rangle$ with $F=F_{\max}\equiv I+3/2$, which is
not disturbed by the control mode, this function is simply defined
\begin{equation}
G_{nn'}^{(--)}(E)=\delta_{nn'}\frac{\hbar}{E-E_n+i\hbar\gamma/2}
\label{a.3}
\end{equation}
where $E_n$ is the energy of the state and $\gamma$ is its natural
relaxation rate. The Green's functions for two other contributing
states $|n\rangle,|n'\rangle$ form a matrix block of $2\times 2$
(with $F=I+1/2,I-1/2$ and with same $M$). However, these atomic
resonances are strongly affected by the interaction with the
control mode such the full set of the coupled master equations
generally includes a block of $3\times 3$ matrix components
$G_{nn'}^{(--)}(E)$ (with $F=I+1/2,I-1/2$ and additionally
$F=I-3/2$ all with same $M$, see Figure \ref{fig6}) and this block
of equations is given by
\begin{equation}
\sum_{n''}\left[\left(E-E_n+i\hbar\frac{\gamma}{2}\right)\delta_{nn''}-
\frac{V_{nm'}V_{n''m'}^{*}}{E-\hbar\omega_c-E_{m'}}\right]G_{n''n'}^{(--)}(E)=\hbar\delta_{nn'}
\label{a.4}
\end{equation}
where $V_{nm'}$ and $V_{n''m'}$ are the matrix elements of interaction with the
control mode. It is important that for each particular block of states
$|n\rangle$, $|n'\rangle$ and $|n''\rangle$ there is only one Zeeman sublevel
in the ground state $|m'\rangle\equiv (F_0=F_{-}=I-1/2,M_0=M)$, which is
coupled with them. The same is also valid for the populated state $|m\rangle$, if
one reconsiders the problem in the basis of circular polarizations in the
$(X,Y)$-plane, as is shown in Figure \ref{fig6}. This basis is given by the
following alternative complete set of unit vectors
\begin{equation}
\mathbf{e}_0=\mathbf{e}_Z, \ \ \ \mathbf{e}_{\pm 1}=\mp\frac{\mathbf{e}_X\pm i\mathbf{e}_Y}{\sqrt{2}}
\label{a.5}
\end{equation}
and it was used in our calculations. Thus for each particular block of states
$|n\rangle$, $|n'\rangle$ and $|n''\rangle$ there is only one pair of the
$\Lambda$-coupled Zeeman sublevels $|m\rangle$ and $|m'\rangle$ in the ground
state and each block of equations (\ref{a.4}) is fully specified by only one
state $|m\rangle$.

The susceptibility tensor in its major frame can be subsequently built via
analytical solution of equations (\ref{a.4}) for each $m$ and next by
substituting the result into expression (\ref{a.2}). Then the tensor's
components in an arbitrary frame can be found via general rotational
transformation. The "dressing" of the atoms by the control mode creates a
quasi-energy resonance structure, which is known as Autler-Townes (AT) effect,
see Refs.\cite{AutlerTownes,LethChebt}, and the location of the AT-resonances
is shown by dashed red bars in Figure \ref{fig6}. Because of the AT effect the
susceptibility tensor becomes anisotropic and can be naturally divided into
linear isotropic and nonlinear anisotropic parts as is shown by
Eq.(\ref{2.2}). The propagation of the signal pulse or its scattered fragment
in any direction can be described by the standard approach of macroscopic
Maxwell theory extended by the Green's function formalism, similar to how it
was done in \cite{DSKH}.

Any probe wave impinging a medium and originally freely propagating in it will
be scattered by atomic dipoles. The scattering process in the medium is
conveniently described by the scattering tensor formalism, see
Ref.\cite{BerestLifshPit}. This tensor is responsible for frequency- and
polarization-dependent transformation of an incident electromagnetic plane wave
as a result of its scattering on an isolated atom. The scattering tensor is
given by
\begin{eqnarray}
\hat{\alpha}_{pq}^{(m''m)}(\omega)%
&\equiv& {\alpha}_{pq}^{(m''m)}(\omega)|m''\rangle\langle m|%
\nonumber\\%
\nonumber\\%
&&\hspace{-0.8 in}%
=-\sum_{n,n'}\frac{1}{\hbar}%
(d_p)_{m''n}(d_q)_{n'm}\,G_{nn'}^{(--)}(\hbar\omega+E_m+i0)\,%
|m''\rangle\langle m|
\label{a.6}%
\end{eqnarray}
where the tensor indices defined either in a Cartesian basis $p,q=x,y,z$ or in
a complex angular momentum basis $p,q=0,\pm 1$ for the atomic dipole operators
can be confined with any reference frame. The most important difference between
expressions (\ref{a.2}) and (\ref{a.6}) is that in the scattering tensor the
output transition is open for any accessible atomic state $|m''\rangle$, i. e.
the entire scattering process includes all the elastic Rayleigh, elastic Raman,
and inelastic Raman scattering channels.

\end{document}